# Transmission Phase of a Quantum Dot with Kondo Correlation Near the Unitary Limit

Yang Ji, M. Heiblum, and Hadas Shtrikman

*Braun Center for Submicron Research, Department of Condensed Matter Physics, Weizmann Institute of Science, Rehovot 76100, Israel*

(November 15, 2018)

The complex transmission amplitude — both magnitude and phase — of a quantum dot (QD) with Kondo correlation was measured near the unitary limit. Contrary to previous phase measurements, performed far from this limit [Ji *et al.*, Science **290**, 779 (2000)], the transmission phase was observed to evolve linearly over a range of about $1.5\pi$ when the Fermi energy was scanned through a *Kondo pair* — a pair of spin degenerate energy levels. Moreover, the phase in Coulomb blockade (CB) peak, adjacent to the Kondo pair, retained a *memory* of the Kondo correlation and did not exhibit the familiar behavior in the CB regime. These results do not agree with theoretical predictions, suggesting that a full explanation may go beyond the framework of the Anderson model.

The Kondo effect, a well known many-body phenomenon arising from the magnetic interaction between a magnetic impurity atom and many free electrons in a host metal, has attracted considerable interest since it is a prime example of a strongly correlated system [1]. Several theoretical groups predicted that the Kondo effect could also be observed in a spin polarized quantum dot (QD) strongly coupled to electron reservoirs [2], which can be described by the Anderson model [3]. Goldhaber-Gordon *et al.* [4] realized the first tunable Kondo effect in such QD, with easy control of the most relevant parameters such as, the energy of the quantized state in the QD and the coupling strength of the QD to the leads. While most Kondo correlated systems have been studied via conductance measurements [4,5], the issue of coherence and phase evolution was neglected until recently [6]. Theoretical prediction [7] for the scattering phase of an electron scattering off a Kondo cloud was found $\pi/2$, independent of the energy of the localized state of the magnetic impurity. This is a consequence of the Kondo-enhanced, Lorentzion type, density of states that is pinned at the Fermi level in the leads. Electrons at the Fermi level, being always at the peak of the Kondo resonance, acquire a constant phase shift of $\pi/2$. For a tunable QD, the transmission amplitude's magnitude and phase evolve as the pair of spin degenerate energy levels in the QD are being scanned through the Fermi level in the leads. Gerland *et al.* [8] predicted the phase to evolve by $\pi$ when such a Kondo pair is being scanned, with a wide plateau of $\pi/2$ throughout the conductance valley (Kondo Valley) that separates the Kondo Pair. Indeed, our recently measured phase showed such a trend , but the phase evolved over a span twice lager [6]. Note, however, that the previously measured QDs were weakly correlated, casting some doubts on the applicability of the conclusions to a strongly correlated system. Here we show results of transmission phase in a strongly correlated QD, in the so-called *unitary limit*, and find even more peculiar and unexpected behavior.

We start with a short description of the system under study. A QD is a small, confined, puddle of electrons coupled to electron reservoirs via tunnel junctions. Its small capacitance ($\sim 10^{-16}F$) leads to a large charging energy, $U_C$, required to add a single electron to the QD. At low enough temperature ($k_B T << U_C$), this results with the appearance of almost periodic conductance peaks, as a function of an externally applied potential, separated by almost zero conductance valleys. This is the well known *Coulomb blockade* (CB) phenomenon [9]. When the top-most spin-degenerate energy level is singly occupied, the QD, which has a non-zero net spin, acts like a localized magnetic impurity. When the unpaired electron in the QD is well coupled to the electron reservoirs, its spin is screened by opposite spin free electrons, creating a dynamic spin-singlet at temperatures lower than the binding energy of the spin singlet — the Kondo temperature $T_K$. This dynamic spin correlation leads to an enhanced density of states centered at the Fermi level (see Fig. 1a), fundamentally altering the properties of the system [10]. Most profoundly, the conductance in the Kondo valley (when the QD has a non-zero net spin), is markedly enhanced, reaching $2e^2/h$ ($e$ is electron charge, $h$ is Planck constant) at the unitary limit [11]. The enhanced conductance can be easily quenched by increasing the temperature, applying a finite DC bias across the QD, or diminishing the coupling strength to the leads [4–6]. While the conductance measurement of a system directly reflects the magnitude of its transmission amplitude, it does not give any information on the coherent nature and phase of the system. These can be obtained, for example, by invoking an electronic two path interferometer with a QD embedded in one of its two paths (Fig. 1b) [12,13,6]. Such structure was formed by negative biasing of sub micron metallic gates laid on the surface of a GaAs-AlGaAs heterostructure with a high mobility two dimensional electron gas (2DEG) embedded $55nm$ below the surface (density $n = 3 \times 10^{11} cm^{-2}$, mobility $\mu = 5 \times 10^5 cm^2 V^{-1} s^{-1}$, measured at $1.5K$). One finds in Fig. 1b three different regions: source (S), drain (D), and a few base regions (B). The base regions are grounded, collecting the back-scattered electrons to ensure that only the two forward-propagating paths (dashed lines in Fig. 1b) reach the drain. In the left arm a tiny QD ($180nm \times 200nm$) is embedded, with both of its *quantum point contacts* (QPCs) and the *plunger* gate, P, individually controlled. The plunger gate is used to tune the potential in the QD, thus controlling the number of electrons in the dot. The right arm provides a reference path to enable two-path interference in the drain. The QD has a charging energy $U_C \sim 1.5 meV$ and a relatively large energy level spacing $\Delta \sim 0.5 meV$, allowing strong coupling to the leads without overlapping of energy levels. A *barrier* gate is added in order to shut off the reference arm and to allow testing of the bare QD. The drain current depends on the complex transmission amplitude of the QD, with magnitude, $t_{QD}$, and phase, $\phi_{QD}$, with $t_{ref}$ and $\phi_{ref}$ belonging to the reference arm. Since $t_{SD} = t_{ref} + t_{QD}$ (assuming $t_{left} = t_{QD}$), the collected current in the drain is $I_{SD} \propto |t_{SD}|^2 = |t_{left}|^2 + |t_{QD}|^2 + 2|t_{left}||t_{QD}|cos(\phi_{ref} - \phi_{QD})$.

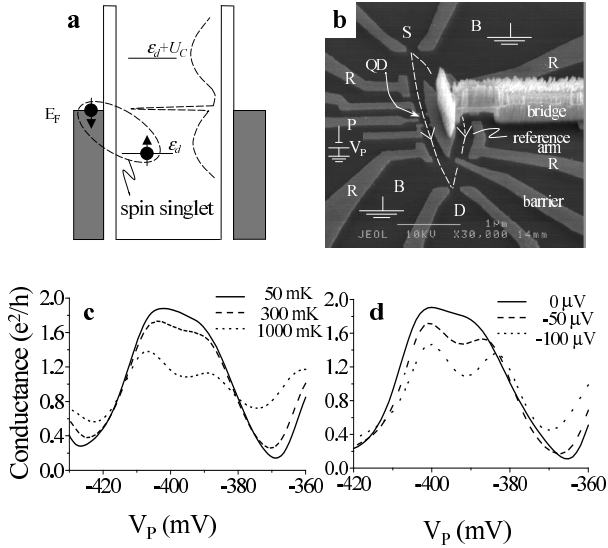

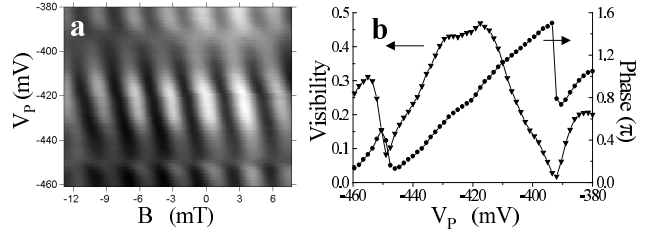

FIG. 2. a. 2D plot of drain current as a function of $V_P$ and $B$. Bright means high current, dark means low current. b. Magnitude (proportional to the visibility of the AB oscillation) and phase of the transmission amplitude of the QD tuned to nearly the unitary limit of the Kondo effect.

FIG. 1. a. Energy scheme of a Kondo correlated QD. The spin degenerate energy levels interferometer. The light pray regions are metallic gates. Note the metallic bridge that biases the central island without crossing the reight arm. c. Differential conductance of the QD as a function of $V_P$ at different temperatures. d. Differential conductance at different $V_{DC}$ across the QD. In both cases valley enhancement quenchs and the Kondo pair is resolved.

Introducing a magnetic flux, $\Phi$, in the area encompassed by the two paths, changes the relative phase of the reference arm via the Aharonov-Bohm (AB) effect [12,13], $\phi_{ref} \longrightarrow \phi_{ref} + 2\pi\Phi/\Phi_0$, where $\Phi_0 = h/e$ is the flux quantum, leading to an oscillating periodic component in the current as a function of magnetic field $\propto cos(\phi_{ref} - \phi_{QD} + 2\pi\Phi/\Phi_0)$. The transmission phase $\phi_{QD}$ can be directly extracted from the phase of the periodic current oscillations. All measurements were done in a dilution refrigerator with temperature $T_{refrigerator} \approx 10mK$ and electron temperature $T_{electron} \approx 50mK$, with an excitation voltage $10\mu V$ oscillating at $7Hz$.

We first identified Kondo correlation by tuning the bare QD and measuring its conductance (after pinching off the reference arm with the barrier gate). A strong enhancement of valley conductance between two adjacent conductance peaks is seen in Figs. 1c and 1d. A peak conductance of $\sim 1.9e^2/h$ was measured, suggesting that the QD is (almost) in the unitary limit. Note that the two low conductance valleys, just before and just after the Kondo pair, with (presumably) zero net spin in the QD, are Coulomb blockaded — as expected. As we increased the temperature (Fig. 1c) or the DC bias across the QD (Fig. 1d), a clear valley was formed and the single broad peak dissolved into two distinct peaks. However, the conductance of the two outer CB valleys increased [4,11]. This is the typical behavior of the conductance in the region of a Kondo pair.

Having identified the Kondo pair, we removed the barrier gate voltage and formed the source (S) and drain (D) QPCs of the interferometer (see Fig. 1b), thus allowing two path interference to take place. The drain current as function of both plunger gate voltage, $V_P$, and magnetic field, $B$, applied perpendicular to the 2DEG, is shown in the gray scale 2D plot in Fig. 2a. Clear AB oscillations, with period $\sim 3.5mT$, and strong phase dependence on $V_P$ are seen. It is easy to notice the abrupt phase slip around $V_P = -450mV$ and $-390mV$. The average visibility, however, is directly related to the magnitude of the coherent transmission amplitude. The visibility and phase, as functions of $V_P$, results are summarized in Fig. 2b. The similarity between the visibility and the conductance indicates that transport is mostly coherent. Moreover, contrary to previous measurements [6], the transmission phase increases almost linearly and spans $\sim 1.5\pi$ within the range where the magnitude of the transmission amplitude is almost constant. Note that the phase minimum in the CB valley at $V_P = -390 mV$ differs by some $0.6\pi$ from that in the CB valley at $V_P = -450 mV$. This is also quite different from the familiar behavior in the CB regime [13].

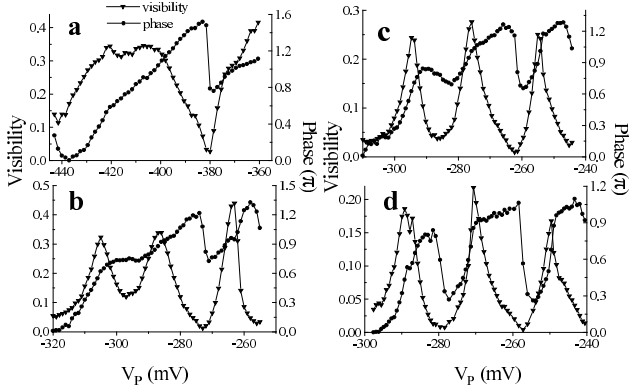
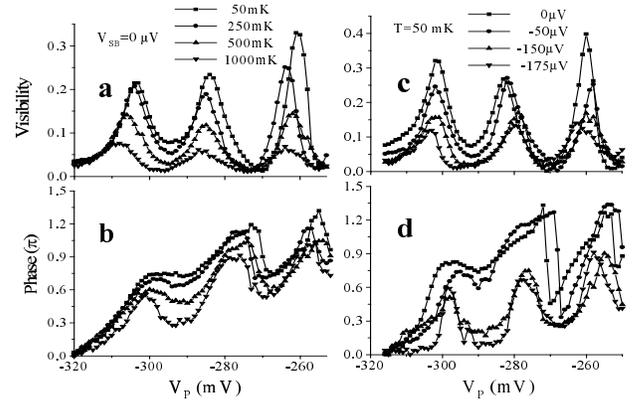

FIG. 3. The dependence of the complex transmission amplitude (magnitude and phase) on the coupling strength of the QD to the leads. The coupling gets weaker from a to d, and the QD moves from Kondo regime to CB regime.

As we pinch off the two QPCs that form the QD, we expect the Kondo correlation to cease and the valley enhancement to quench. Figure 3 shows the visibility and the phase as we add three electrons to the QD, namely, as we scan through a Kondo pair, and an adjacent CB related peak. Indeed, as we reduced the coupling strength, the broad peak of the visibility developed a valley and split into two separated peaks — in accordance with the conductance measurement [4–6,11]. The phase evolution, however, which climbed almost linearly by some $1.5\pi$ in the unitary limit, developed a plateau and later, as the coupling weakened further and the QD entered the CB regime, exhibited a phase lapse (Fig. 3a through 3d). The total phase shift through the Kondo pair, which spanned $\sim 1.5\pi$ in the unitary limit, changed to the familiar span of $\sim \pi$ in the CB regime, with an almost periodic phase behavior [14]. Note the striking phase behavior outside the Kondo pair. While the very sharp phase slip that follows the Kondo pair persists at all coupling strengths, the absolute value of the phase in the CB valley, after slipping (near $V_P = -380 mV$), was $\sim 0.8\pi$ when the QD was in the unitary limit and only $\sim 0.2\pi$ (near $V_P = -255 mV$) when the QD was pinched off to the CB regime. More clearly stated, a quench of Kondo correlation affects the phase in the CB valley and that of the conductance peak

FIG. 4. a. Magnitude and b. phase of the transmission amplitude of the QD measured at different temperatures. c. Magnitude and d. phase of the transmission amplitude measured at different DC bias applied between the source and the base regions. Both temperature and DC bias quench the conductance enhancement in the Kondo valley.

that is adjacent to the Kondo pair.

Similarly, increasing the temperature to the order of $T_K$ or increasing the energy of the impinging electrons (by applying $V_{SB}$) to around $k_B T_K$ is expected to destroy the Kondo correlation (see Fig. 1). Figure 4 shows the complex transmission amplitude at different temperatures and DC bias $V_{SB}$. Note that the dephasing length in the interferometer drops with increasing temperature and energy (leading to a reduced visibility), we were limited to $T < 1K$ and $V_{SB} < -200\mu V$. Consequently, we had to reduce the Kondo temperature to $T_K \sim 1.5K$ by somewhat pinching off the QD in order to observe an effect. Then, when the temperature increased (Figs. 4a and 4b), the visibility followed the behavior of the conductance, but the phase evolution changed from that with a plateau of $\sim 0.8\pi$ in the Kondo valley [6] to a phase lapse at high temperatures. We attribute the fact that the phase lapse did not reach a full $-\pi$ lapse even at $1K$ to the still relatively high Kondo temperature. Similarly, applying a small DC bias to the source at the lowest temperature leads to a similar change in the phase evolution, moving from a smooth increase with a plateau for $V_{SB} = 0$ to a phase lapse for $V_{SB} = -150\mu V$ (Fig. 4d). Again, in both cases, the phase slip in the adjacent CB valley ($V_P = -270 mV$) moved down rigidly with the phase change in the Kondo pair as the correlation was quenched.

While the phase behavior in the CB regime is familiar by now [13], the behavior when Kondo correlation sets in — be it at low enough temperature or when the coupling to the leads is very strong — is puzzling. Two main (troubling) features stand out. The first is the peculiar behavior of the phase and its large span — twice larger than the predicted value [8]. Recall that in earlier experiments [6] the temperature was a bit higher ($\sim 100 mK$) and the coupling to the leads was weaker, both leading to a phase span throughout the Kondo pair of $\sim 2\pi$ with a clear plateau of $\pi$ throughout in the Kondo enhanced valley. Here, however, the electrons' temperature is lower ($\sim 50 mK$) and the coupling to the leads is stronger, resulting with a larger $T_K$. Hence, a full blown enhancement of the valley conductance and an almost linear phase rise of $\sim 1.5\pi$ through the Kondo pair are observed. One may hypothesize that the linear phase rise comes as a result of the added phase contributions of both spin degenerate, relatively broad, single particle levels and the Kondo resonance centered at the Fermi surface (Fig. 1a). These added phase contributions can, under some conditions, indeed eliminate the $\pi/2$ plateau, as found in a numerical example in Ref. 8. However, in that calculated example the span of the predicted phase rise was always smaller than $\pi$. The large phase span observed in our experiments contradicts the above hypothesis. The second striking feature is the phase behavior adjacnet to the Kondo pair. A naive expectation, based on the Anderson model [10], is that Kondo correlation affects only the property of the Kondo valley, when the QD has non zero net spin. When the spin degenerate level is doubly occupied though (Fig. 1c), the QD should exhibit standard CB behavior with no *memory* of the spin correlation. In other words, the adjacent CB conductance valley should be low and the phase there should be the characteristic phase in the CB regime (namely, return to zero). However, our results clearly show that Kondo correlation dramatically affects the phase in the adjacent, non-Kondo, CB valley. And more surprisingly, as the correlation is being destroyed (say, via a weaker coupling, higher temperature, or an applied voltage), the phase behavior in the adjacent non-Kondo valley alters and returns to its its characteristic behavior in the CB regime. This means that the QD, somehow, *remembers* the occurrence of Kondo correlation even after it ceases to exist. An explanation for the puzzling phase behavior may go beyond the simple Anderson model.

*Acknowledgement* We are grateful to D. Mahalu for doing the ebeam writing of the complex structure. We thank D. Goldhaber-Gordon, E. Buks, D. Sprinzak, and Y. Chung for helpful information on sample fabrication and measuring techniques. We also thank Y. Oreg, J. von Delft and P. Brouwer for stimulating discussions. The work was partly supported by the MINERVA foundation and by German Israeli Project Cooperation (DIP).